\documentstyle[twoside,psfig]{article}
\newcommand{\be}{\begin{equation}}
\newcommand{\ee}{\end{equation}}
\newcommand{\beq}{\begin{eqnarray}}
\newcommand{\eeq}{\end{eqnarray}}

\newcommand{\lie}[1]{\mbox{$:\! #1 \!:$}}
\mathchardef\calM="224D
\mathchardef\calP="2250

\catcode`\@=11
\long\def\@makefntext#1{
\protect\noindent \hbox to 3.2pt {\hskip-.9pt
$^{{\eightrm\@thefnmark}}$\hfil}#1\hfill}       %CAN BE USED

\def\@makefnmark{\hbox to 0pt{$^{\@thefnmark}$\hss}}    %ORIGINAL

\def\ps@myheadings{\let\@mkboth\@gobbletwo
\def\@oddhead{\hbox{}
\rightmark\hfil\eightrm\thepage}
\def\@oddfoot{}\def\@evenhead{\eightrm\thepage\hfil
\leftmark\hbox{}}\def\@evenfoot{}
\def\sectionmark##1{}\def\subsectionmark##1{}}

\oddsidemargin=\evensidemargin
\newcounter{sectionc}\newcounter{subsectionc}\newcounter{subsubsectionc}
\renewcommand{\section}[1] {\vspace{12pt}\addtocounter{sectionc}{1}
\setcounter{subsectionc}{0}\setcounter{subsubsectionc}{0}\noindent
    {\tenbf\thesectionc. #1}\par\vspace{5pt}}
\renewcommand{\subsection}[1] {\vspace{12pt}\addtocounter{subsectionc}{1}
    \setcounter{subsubsectionc}{0}\noindent
    {\bf\thesectionc.\thesubsectionc. {\kern1pt \bfit #1}}\par\vspace{5pt}}
\renewcommand{\subsubsection}[1] {\vspace{12pt}\addtocounter{subsubsectionc}{1}
    \noindent{\tenrm\thesectionc.\thesubsectionc.\thesubsubsectionc.
    {\kern1pt \tenit #1}}\par\vspace{5pt}}
\newcommand{\nonumsection}[1] {\vspace{12pt}\noindent{\tenbf #1}
    \par\vspace{5pt}}

\newcounter{appendixc}
\newcounter{subappendixc}[appendixc]
\newcounter{subsubappendixc}[subappendixc]
\renewcommand{\thesubappendixc}{\Alph{appendixc}.\arabic{subappendixc}}
\renewcommand{\thesubsubappendixc}
    {\Alph{appendixc}.\arabic{subappendixc}.\arabic{subsubappendixc}}

\renewcommand{\appendix}[1] {\vspace{12pt}
        \refstepcounter{appendixc}
        \setcounter{figure}{0}
        \setcounter{table}{0}
        \setcounter{lemma}{0}
        \setcounter{theorem}{0}
        \setcounter{corollary}{0}
        \setcounter{definition}{0}
        \setcounter{equation}{0}
        \renewcommand{\thefigure}{\Alph{appendixc}.\arabic{figure}}
        \renewcommand{\thetable}{\Alph{appendixc}.\arabic{table}}
        \renewcommand{\theappendixc}{\Alph{appendixc}}
        \renewcommand{\thelemma}{\Alph{appendixc}.\arabic{lemma}}
        \renewcommand{\thetheorem}{\Alph{appendixc}.\arabic{theorem}}
        \renewcommand{\thedefinition}{\Alph{appendixc}.\arabic{definition}}
        \renewcommand{\thecorollary}{\Alph{appendixc}.\arabic{corollary}}
        \noindent{\tenbf Appendix \theappendixc #1}\par\vspace{5pt}}
\newcommand{\subappendix}[1] {\vspace{12pt}
        \refstepcounter{subappendixc}
        \noindent{\bf Appendix \thesubappendixc. {\kern1pt \bfit #1}}
    \par\vspace{5pt}}
\newcommand{\subsubappendix}[1] {\vspace{12pt}
        \refstepcounter{subsubappendixc}
        \noindent{\rm Appendix \thesubsubappendixc. {\kern1pt \tenit #1}}
    \par\vspace{5pt}}

\newcommand{\textlineskip}{\baselineskip=13pt}
\newcommand{\smalllineskip}{\baselineskip=10pt}

\def\eightcirc{
\begin{picture}(0,0)
\put(4.4,1.8){\circle{6.5}}
\end{picture}}
\def\eightcopyright{\eightcirc\kern2.7pt\hbox{\eightrm c}}

\def\abstracts#1#2#3{{
    \centering{\begin{minipage}{4.5in}\footnotesize\baselineskip=10pt
    \parindent=0pt #1\par
    \parindent=15pt #2\par
    \parindent=15pt #3
    \end{minipage}}\par}}

\def\keywords#1{{
    \centering{\begin{minipage}{4.5in}\footnotesize\baselineskip=10pt
    {\footnotesize\it Keywords}\/: #1
    \end{minipage}}\par}}

\renewenvironment{thebibliography}[1]
        {\frenchspacing
     \ninerm\baselineskip=11pt
         \begin{list}{\arabic{enumi}.}
        {\usecounter{enumi}\setlength{\parsep}{0pt}
     \setlength{\leftmargin 12.7pt}{\rightmargin 0pt} %FOR 1--9 ITEMS
         \setlength{\itemsep}{0pt} \settowidth
    {\labelwidth}{#1.}\sloppy}}{\end{list}}

\newcounter{itemlistc}
\newcounter{romanlistc}
\newcounter{alphlistc}
\newcounter{arabiclistc}

\newcommand{\fcaption}[1]{
        \refstepcounter{figure}
        \setbox\@tempboxa = \hbox{\footnotesize Fig.~\thefigure. #1}
        \ifdim \wd\@tempboxa > 5in
           {\begin{center}
        \parbox{5in}{\footnotesize\smalllineskip Fig.~\thefigure. #1}
            \end{center}}
        \else
             {\begin{center}
             {\footnotesize Fig.~\thefigure. #1}
              \end{center}}
        \fi}

\newcommand{\tcaption}[1]{
        \refstepcounter{table}
        \setbox\@tempboxa = \hbox{\footnotesize Table~\thetable. #1}
        \ifdim \wd\@tempboxa > 5in
           {\begin{center}
        \parbox{5in}{\footnotesize\smalllineskip Table~\thetable. #1}
            \end{center}}
        \else
             {\begin{center}
             {\footnotesize Table~\thetable. #1}
              \end{center}}
        \fi}

\def\@citex[#1]#2{\if@filesw\immediate\write\@auxout
    {\string\citation{#2}}\fi
\def\@citea{}\@cite{\@for\@citeb:=#2\do
    {\@citea\def\@citea{,}\@ifundefined
    {b@\@citeb}{{\bf ?}\@warning
    {Citation `\@citeb' on page \thepage \space undefined}}
    {\csname b@\@citeb\endcsname}}}{#1}}

\newif\if@cghi
\def\cite{\@cghitrue\@ifnextchar [{\@tempswatrue
    \@citex}{\@tempswafalse\@citex[]}}
\def\citelow{\@cghifalse\@ifnextchar [{\@tempswatrue
    \@citex}{\@tempswafalse\@citex[]}}
\def\@cite#1#2{{$\null^{#1}$\if@tempswa\typeout
    {IJCGA warning: optional citation argument
    ignored: `#2'} \fi}}

\def\pmb#1{\setbox0=\hbox{#1}
    \kern-.025em\copy0\kern-\wd0
    \kern.05em\copy0\kern-\wd0
    \kern-.025em\raise.0433em\box0}

\def\fnt#1#2{\footnotetext{\kern-.3em
    {$^{\mbox{\scriptsize #1}}$}{#2}}}

\def\ps@myheadings{%
    \let\@oddfoot\@empty\let\@evenfoot\@empty
    \def\@evenhead{\slshape\leftmark\hfil}%       %EVEN PAGE
    \def\@oddhead{\hfil{\slshape\rightmark}}%     %ODD PAGE
    \let\@mkboth\@gobbletwo
    \let\sectionmark\@gobble
    \let\subsectionmark\@gobble
    }
\font\tenrm=cmr10
\font\tenit=cmti10
\font\tenbf=cmbx10
\font\bfit=cmbxti10 at 10pt
\font\ninerm=cmr9

\font\eightrm=cmr8

\def\qed{\hbox{${\vcenter{\vbox{            %HOLLOW SQUARE
   \hrule height 0.4pt\hbox{\vrule width 0.4pt height 6pt
   \kern5pt\vrule width 0.4pt}\hrule height 0.4pt}}}$}}

\def\bsc{{\sc a\kern-6.4pt\sc a\kern-6.4pt\sc a}}   %LATEX LOGO
\def\bflatex{\bf L\kern-.30em\raise.3ex\hbox{\bsc}\kern-.14em
T\kern-.1667em\lower.7ex\hbox{E}\kern-.125em X}

\begin{document}
\setlength{\textheight}{7.7truein}  %for 2nd page onwards

\thispagestyle{empty}

\markboth{\protect{\footnotesize\it Polynomial Map Symplectic Algorithm
}}{\protect{\footnotesize\it Polynomial Map Symplectic Algorithm
}}

\normalsize\textlineskip

\setcounter{page}{1}

%\copyrightheading{}         %{Vol. 0, No. 0 (1993) 000--000}

\vspace*{0.88truein}

%\fpage{1}
\centerline{\bf POLYNOMIAL MAP SYMPLECTIC ALGORITHM}
\vspace*{0.37truein}
\centerline{\footnotesize GOVINDAN RANGARAJAN}
\baselineskip=12pt
\centerline{\footnotesize\it Department of Mathematics and Centre for Theoretical Studies,
Indian Institute of Science}
\baselineskip=10pt
\centerline{\footnotesize\it Bangalore 560 012, India}
\centerline{\footnotesize\it E-mail: rangaraj@math.iisc.ernet.in}

%\vspace*{10pt}     %actual spacing
%\vspace*{0.225truein}
%\publisher{(received date)}{(revised date)}

\vspace*{0.25truein}
\abstracts{Long-term stability studies of
nonlinear Hamiltonian systems require symplectic integration
algorithms which are both fast and accurate. In this paper, we study
a symplectic integration method wherein the symplectic
map representing the Hamiltonian system is refactorized using polynomial
symplectic maps. This method is analyzed in detail
for the three degree of freedom case. We obtain explicit
formulas for the action of the constituent polynomial maps on
phase space variables.}{}{}

\vspace*{5pt}

%\textlineskip          %) USE THIS MEASUREMENT WHEN THERE IS
%\vspace*{12pt}         %) NO SECTION HEADING

\vspace*{1pt}\textlineskip  %) USE THIS MEASUREMENT WHEN THERE IS
\section{Introduction}     %) A SECTION HEADING
\vspace*{-0.5pt}
\noindent
Numerical integration algorithms are essential to study the long term
single particle stability of nonlinear, nonintegrable Hamiltonian
systems.  However,
standard numerical integration algorithms can not be used since they are not
symplectic \cite{sanzserna}. This violation of the symplectic condition can lead to spurious
chaotic or dissipative behavior. Numerical integration algorithms which
satisfy the symplectic condition are called symplectic integration algorithms
\cite{sanzserna}.
Several symplectic integration algorithms have been proposed in the literature
\cite{ruth,kang,lasagni,neri,irwin,suris,yoshida,forest,channell,gr1,gr2,dragt1,iserles,okubnor,sanzserna2,dragt2,gjaja,gr3,gr4,gr5}.
Some of these directly use the Hamiltonian whereas others use
the symplectic map \cite{Dragt1,Dragt2} representing the nonlinear Hamiltonian system.
For complicated systems like the Large Hadron Collider which has
thousands of elements, using individual Hamiltonians for each element can drastically
slow down the integration process. One the other hand, the map based approach
is very fast in such cases \cite{yan,berg}.

One class of the map-based methods uses jolt
factorization \cite{irwin,gr1,dragt2,gr3}. But there are still unanswered questions on how to
best choose the underlying group and elements in the group \cite{etienne}.
Further, some of these methods \cite{gr1,dragt2,gr3} can be quite difficult to generalize to
higher dimensions. Another class of methods uses solvable maps \cite{gr2,gr5}
or monomial maps \cite{gjaja}. Even though they are fairly straightforward
to generalize to higher dimensions, they tend to introduce spurious poles and branch
points not present in the original map \cite{etienne}.

We investigate a new symplectic integration method
where the symplectic map is refactorized using ``polynomial maps''
(maps whose action on phase space variables gives rise to
polynomials). This does not introduce spurious poles and branch
points. Moreover, it is easy to generalize to higher dimensions.
Further, since it is map-based, it is also very fast.
In this paper, we describe in detail the theoretical
underpinnings of the polynomial map factorization of symplectic
maps. We also apply it to Hamiltonian systems.

\section{Preliminaries}
\noindent
We restrict ourselves to three degrees of freedom nonlinear Hamiltonian
system.
The effect of a Hamiltonian system on a particle can be formally expressed
as the action of a symplectic map ${\calM}$ that takes the particle from its
initial state $z^{in}$ to its final state $z^{fin}$ \cite{Dragt1,Dragt2}
\be
z^{fin} = {\calM}\, z^{in}.
\label{I.15}\ee
Here $z$ represents the collection of six phase space variables:
\be
z = (q_1,q_2,q_3,p_1,p_2,p_3).
\ee
Using the Dragt-Finn factorization theorem\cite{Finn,Dragt1}, the symplectic map
${\calM}$ can be factorized as shown below:
\be
{\calM} = {\hat M} e^{:f_3:}\,e^{:f_4:}\, \ldots e^{:f_n:} \ldots \ .
\label{2}\ee
Here $f_n(z)$
denotes a homogeneous polynomial (in $z$) of degree $n$ uniquely
determined by the factorization theorem. The Lie transformation
$e^{:f(z):}$ is given by
\be
e^{:f(z):} = \sum_{n=0}^{\infty} \frac{\lie{f(z)}^n}{n!},
\label{1b}\ee
where
\be
\lie{f(z)} g(z) = [f(z),g(z)].
\label{1a}\ee
Here $[f(z),g(z)]$ denotes the usual Poisson bracket of the functions $f(z)$
and $g(z)$.
Further ${\hat M}$ gives the linear part of the map and hence has an
equivalent representation in terms of the Jacobian matrix $M$ of the
map ${\calM}$ \cite{Dragt1}:
\be
{\hat M} z_i = M_{ij} z_j = (Mz)_i.
\label{3}\ee
The infinite product of Lie transformations $\exp(\lie{f_n})$ $(n=3,4, \ldots$)
in Eq. (\ref{2}) represents the nonlinear part of ${\calM}$.

As an application, let us consider a charged particle particle storage ring
which typically comprises thousands of elements (drifts, quadrupoles, sextupoles etc.)
Using the above procedure, one can represent each element in the storage ring
by a symplectic map. By concatenating \cite{Dragt1} these maps together using
group-theoretical methods \cite{Cornwell}, we obtain the
so-called `one-turn' map representing the entire storage ring.
The one-turn map
gives the final state $z^{(1)}$ of a particle after one turn around the ring as a
function of its initial state $z^{(0)}$:
\be
z^{(1)} = {\calM} z^{(0)}.
\ee
To obtain the state of a particle after $n$ turns, one has to merely
iterate the above mapping $N$ times i.e.
\be
z^{(n)} = {\calM}^n z^{(0)}.
\label{rec} \ee
Since ${\calM}$ is explicitly symplectic, this gives a symplectic integration
algorithm. Further, since the entire ring can be represented by a single (or
at most a few) symplectic map(s), numerical integration of particle trajectories using
symplectic maps is very fast.

To obtain a practical symplectic integration algorithm, we follow
the perturbative approach and truncate
${\calM}$ after a finite number of Lie transformations:
\be
{\calM} \approx {\hat M} e^{:f_3:}\,e^{:f_4:}\, \ldots e^{:f_P:}.
\label{4}\ee
The symplectic map is said to be truncated at order $P$.
This map is still symplectic.
However, each exponential $e^{:f_n:}$
in ${\calM}$ still contains an infinite number of terms in its Taylor
series expansion.
We get around the above problem by refactorizing ${\calM}$ in terms
of simpler symplectic maps which can be evaluated exactly without truncation.
We use `polynomial maps' which give rise to polynomials when acting on the
phase space variables. This avoids the problem of spurious poles and
branch points present in generating function methods \cite{etienne},
solvable map \cite{gr2,gr5} and monomial map \cite{gjaja} refactorizations.

\section{Symplectic Polynomial Maps}
\noindent
In this section we study symplectic polynomial maps in some detail. We start
by describing the difference between monomial maps and
polynomial maps with respect to presence of poles and branch points.
This difference can be illustrated using the following examples.
Consider the monomial symplectic map $\exp(:q_1^2 p_1:)$. Its
action on $q_1$, $p_1$ in a two dimensional
phase space is given as follows:
\be
q_1' = \exp(:q_1^2 p_1:) q_1 = \frac{q_1}{1+q_1}; \ \ \
p_1' = \exp(:q_1^2 p_1:) p_1 = p_1 (1+q_1)^2.
\ee
This map has a pole at $q_1=-1$.

On the other hand, consider the symplectic map $\exp(:a_1 q_1^3 + a_2 p_1:)$
where $a_1$, $a_2$ are real constants. We determine its action
on phase space variables as follows. Note that the symplectic map
is of the form $\exp(:h(z):)$ where $h(z)$ is a function
which depends only on the phase space variables $z$ and is independent
of time $t$. If we take $h(z)$ to be the Hamiltonian function, then solving the
Hamilton's equations of motion for this Hamiltonian from time
$t=t^i$ to time $t=t^f$ is equivalent to the following symplectic
map action \cite{Dragt1}:
\be
z(t=t^f) = \exp[-(t^f-t^i) :h(z):] z(t=t^i).
\ee
Equivalently, obtaining the action of the symplectic map
$\exp[-(t^f-t^i) :h(z):]$ on the phase space variables is
the same as solving the Hamilton's equations of motion with
$h(z)$ as the Hamiltonian from time $t^i$ to $t^f$. Setting
$t^i=0$ and $t^f=-1$ we have the following equivalence: Obtaining
the action of the symplectic map $\exp(:h(z:)$ on phase space
variables is equivalent to solving the Hamilton's equations
of motion using $h(z)$ as the Hamiltonian from time $t=0$
to time $t=-1$. In this case, $z(0)$ will correspond to the
initial values of the phase space variables and $z(-1)$ to
the final values obtained after the action of the map $\exp(:h(z:)$.

Returning to our symplectic map, we obtain its action by first
solving the Hamilton's equations of motion from time $t=0$ to
$t=-1$ using the argument of the Lie transformation, $h = a_1 q_1^3 + a_2 p_1$,
as the Hamiltonian. The Hamilton's equations of
motion are given by:
\beq \label{Heqns}
\frac{dq_1}{dt} & = & \frac{\partial h}{\partial p_1}, \nonumber \\
\frac{dp_1}{dt} & = & -\frac{\partial h}{\partial q_1}.
\eeq
Solving these simple equations, we obtain:
\beq
q_1(t) & = & q_1(0) + a_2 t; \nonumber \\
p_1(t) & = & p_1(0) - a_1 a_2^2 t^3 - 3 a_1 a_2 q_1(0) t^2
- 3 a_1 q_1(0)^2 t.
\eeq
where $q_1(0)$ and $p_1(0)$ denote the values of $q_1$ and $p_1$
at time $t=0$. To obtain the action of the map
$\exp(:a_1 q_1^3 + a_2 p_1:)$ on the phase space variables, we
set $t=-1$ in the above equations and denote $q_1(-1)$, $p_1(-1)$
by $q_1^{fin}$, $p_1^{fin}$ and $q_1(0)$, $p_1(0)$
by $q_1^{in}$, $p_1^{in}$ respectively.  Thus we get
\be
q_1^{fin} = q_1^{in} - a_2, \ \ \ p_1^{fin} = p_1^{in} + a_1 a_2^2 -
3 a_1 a_2 q_1^{in} + 3 a_1 (q_1^{in})^2.
\ee
Using Eq. (\ref{1b}), we can easily verify that the above result is
indeed correct. We note that the final values of the phase space
variables are polynomial functions of the initial variables and
therefore involve no poles or branch points.
This is an example of a polynomial map.

We now determine the classes of symplectic maps which are also
polynomial maps.
We obtain the following simple principles which are equally applicable in
higher dimensions.
\begin{enumerate}
\item{} All polynomials of the form $h(z)$ where both
a phase space variable and its
canonically conjugate variable \cite{Goldstein} do not occur simultaneously
give rise to symplectic polynomial maps via $\exp(:h(z):)$. We will
call such $h(z)$'s as polynomials of the first type.
\item{} If a canonically conjugate pair $q_i,p_i$ is present in the polynomial
$h(z)$ and it appears either in the form $[a(\bar z) q_i+g(p_i,\bar z)]^m$ or
$[a(\bar z) p_i + g(q_i,\bar z)]^m$ (where $m=1,2 \ldots$, $\bar z = \{ q_j,p_k \}$
with $j \ne k \ne i$ and $a$, $g$ are polynomials in the indicated variables),
then this polynomial $h(z)$ again gives rise
to a symplectic polynomial map via $\exp(:h(z):)$. If a product/sum of such factors appears
in $h(z)$, each term in the product/sum is a function of different canonically
conjugate pairs. We will
call $h(z)$'s of the form described above as polynomials of the second type.
\end{enumerate}

We can prove the above results as follows. Let $\hat{z}$ denote a collection of
phase space variables $\{ q_j,p_k \}$ with $j \ne k$. Thus polynomials
of the first type are of the form $h(\hat{z})$. The polynomial map
is then given by $\exp(:h(\hat{z}):)$. As described earlier, its action on the
phase space variables
is given by solving the Hamilton's equations of motion from time $t=0$ to
$t=-1$ using $h(\hat{z})$ as the Hamiltonian. From classical mechanics \cite{Goldstein}
each of the variables in the collection $\hat{z}$ is a cyclic variable and
is therefore conserved by the Hamiltonian. Thus $\hat{z}(t) = \hat{z}(0)$.
Consequently
\be \label{case1a}
\hat{z}^{fin} = \exp(:h(\hat{z}):) \hat{z}^{in} = \hat{z}^{in}.
\ee
Next we consider the action of this map on the variable $\tilde{z_j}$
which is canonically conjugate to $\hat{z}_j$. Solving Hamilton's equations
of motion with $h(\hat{z})$ as the Hamiltonian from time $t=0$ to $t=-1$ we get
\be
\tilde{z_j}(-1) = \tilde{z_j}(0)-(-1)^r \frac{d h}{d \hat{z}_j} (\hat{z}_j(0)),
\ee
where $r$ is zero if $\tilde{z_i}$ is a coordinate variable and 1 otherwise.
As before, the action of $\exp(:h(\hat{z}):)$ on $\tilde{z_j}$ is obtained
by setting $\tilde{z_j}(0)$ as $\tilde{z_j}^{in}$ and $\tilde{z_j}(-1)$
as $\tilde{z_i}^{fin}$:
\be \label{case1b}
\tilde{z_j}^{fin} = \tilde{z_j}^{in} - (-1)^r \frac{d h}{d \hat{z}_j} (\hat{z}_j^{in}).
\ee
Since $h(\hat{z})$ is a polynomial in $\hat{z}$, the right hand side of the above
equation is a polynomial in $\hat{z}$ and $\tilde{z}$.
From Eqs. (\ref{case1a}) and (\ref{case1b}) we conclude that
$\exp(:h(\hat{z}):)$ is a polynomial map. Further, since all Lie
transformations are symplectic maps \cite{Dragt1}, $\exp(:h(\hat{z}):)$
is a symplectic polynomial map.

Next we consider polynomials $h(z)$ of the second type described in item 2) above.
Let $h(z) = [a(\bar z) q_i+g(p_i,\bar z)]^m$ where $q_i,p_i$ is a
canonically conjugate pair. Further $m$ is a positive integer, $\bar z = \{ q_j,p_k \}$
with $j \ne k \ne i$ and $a$, $g$ are polynomials in the indicated variables.
We will show that $\exp(:h(z):)$ is a polynomial map. The proof for
the case $h'(z)=[a(\bar z) p_i + g(q_i,\bar z)]^m$ is similar.
A concrete example of the type of $h(z)$ that we are considering is given
by $h(z) = (\alpha_1 p_2 + \alpha_2 p_3 +
\alpha_3 p_2^2 + \alpha_4 p_2 p_3 + \alpha_5 p_3^2 + \alpha_6 q_3 p_2)^m$
where $\alpha_i$'s are real constants.

As before we first solve the Hamilton's equations of motion with
$h(z)$ as the Hamiltonian. Since each of the variables in the collection
denoted by $\bar z$ is cyclic, we have $\bar{z}(t) = \bar{z}(0)$ and
\be \label{case2a}
\bar{z}^{fin} = \exp(:h(\bar{z}):) \bar{z}^{in} = \bar{z}^{in}.
\ee
Solving the Hamilton's equation of motion for $p_i$ we get
\be
\frac{d p_i}{dt} = - \frac{\partial h(z)}{\partial q_i}
= - m [a(\bar z(0)) q_i(0)+g(p_i(0),\bar z(0))]^{m-1} a(\bar z(0)),
\ee
where we have used the fact that $[a(\bar z) q_i+g(p_i,\bar z)]$
is a conserved quantity under this Hamiltonian flow i.e.
$[a(\bar z) q_i+g(p_i,\bar z)]=[a(\bar z(0)) q_i(0)+g(p_i(0),\bar z(0))]$.
Solving this simple equation we obtain
\be
p_i(t) = p_i(0)
- m [a(\bar z(0)) q_i(0)+g(p_i(0),\bar z(0))]^{m-1} a(\bar z(0)) t.
\ee
Note that $p_i(t)$ is a polynomial in both $z$ and $t$. Setting
$t=-1$ and denoting $z(0),z(-1)$ by $z^{in},z^{fin}$ respectively,
we see that $p_i^{fin} = \exp(:~h(z)~:) p_i^{in}$ is a
polynomial in $z^{in}$:
\be \label{case2b}
p_i^{fin} = p_i^{in}
+ m [a(\bar z^{in}) q_i^{in})+g(p_i^{in},\bar z^{in})]^{m-1} a(\bar z^{in}).
\ee

The equation of motion for $q_i$ gives us
\be
\frac{d q_i}{dt} =  \frac{\partial h(z)}{\partial p_i}
= - m [a(\bar z(0)) q_i(0)+g(p_i(0),\bar z(0))]^{m-1}
\frac{d g(p_i,\bar z)}{d p_i}.
\ee
Since $p_i(t)$ is a polynomial in $z$ and $t$, the right hand
side of the above equation is also a polynomial in $z$ and $t$.
Therefore, the above differential equation can be easily
integrated to give $q_i(t)$ which is guaranteed to be a polynomial
in $z$ and $t$. Setting
$t=-1$ and denoting $z(0),z(-1)$ by $z^{in},z^{fin}$ respectively,
we obtain the result that $q_i^{fin} = \exp(:h(z):) q_i^{in}$ is a
polynomial in $z^{in}$.
Finally, the equation of motion for $\breve z_j$, the variable
canonically conjugate to $\bar z_j$, is given by
\beq
\frac{d \breve{z}_j}{dt} & = & (-1)^r \frac{\partial h(z)}{\partial {\bar z}_j}
= (-1)^r m [a(\bar z(0)) q_i(0)+g(p_i(0),\bar z(0))]^{m-1} \nonumber \\
&& \times \frac{d [a(\bar z) q_i+g(p_i,\bar z)]}{d {\bar z}_j}(\bar z=\bar z(0)),
\eeq
where $r$ is zero if $\breve {z}_j$ is a coordinate variable and 1 otherwise.
The right hand side is a polynomial function of $q_i,p_i$ both of which in turn are
polynomials in $t$. Hence the equation can be integrated giving
$\breve{z}_j(t)$ as a polynomial in $z$ and $t$. Setting $t=-1$
and denoting $z(0),z(-1)$ by $z^{in},z^{fin}$ respectively,
we find that $\breve{z_j}^{fin} = \exp(:h(z):) \breve{z}_j^{in}$ is a
polynomial in $z^{in}$. Thus we have proved that $\exp(:h(z):)$
is a (symplectic) polynomial map.

To conclude, we consider the case where a sum/product of factors
of the form $[a(\bar z) q_i+g(p_i,\bar z)]^m$ or
$[a(\bar z) p_i + g(q_i,\bar z)]^m$ appear in $h(z)$.
If a phase space variable appears in one factor of the sum/product, by
assumption, neither this variable nor its canonically conjugate variable
appears in the remaining factors of the sum/product. Therefore
each term in the sum/product is independently conserved by the
Hamiltonian flow generated by $h(z)$ and acts only on the phase
space variables appearing in that term. Consequently each term can
be considered separately and since each term is of the form
$[a(\bar z) q_i+g(p_i,\bar z)]^m$ or
$[a(\bar z) p_i + g(q_i,\bar z)]^m$, the argument given above
can be immediately applied proving that $\exp(:h(z):)$ is a
symplectic polynomial map even in this case. This completes the proof
of the claims made in items 1) and 2). We conjecture that
all symplectic polynomial maps have one of the two forms enumerated
above.

\section{Symplectic Integration using Polynomial Maps}
\noindent
In this section, we return to the problem of symplectic integration.
We restrict ourselves
to symplectic maps in a six dimensional phase space truncated at order 4.
The results obtained below can be generalized to both higher orders
and higher dimensions using symbolic
manipulation programs. The Dragt-Finn factorization of the symplectic map
is given by:
\be
{\calM} = {\hat M} e^{:f_3:}\,e^{:f_4:},
\ee
where
\beq\label{fs}
f_3 & = & a_{28} q_1^3 + a_{29} q_1^2 p_1 + \cdots + a_{83} p_3^3, \nonumber \\
f_4 & = & a_{84} q_1^4 + a_{85} q_1^3 p_1 + \cdots + a_{209} p_3^4.
\eeq
Here the coefficients $a_{28}, \ldots , a_{209}$ can be explicitly computed given
a Hamiltonian system\cite{Dragt1} and are therefore known to us.
The numbering of these monomial coefficients
follows the standard Giorgilli scheme \cite{giorgilli}.
The above map captures the leading order nonlinearities of the system.
Since the action of the linear part ${\hat M}$ on
phase space variables is well known [cf. Eq. (\ref{3})] and is already a polynomial
action, we only refactorize the nonlinear part of the map using $N$ polynomial maps \cite{grpoly}.
This is done as follows:
\be
{\calM} \approx {\calP} = {\hat M} e^{:h_1:}\,e^{:h_2:} \cdots e^{:h_{N}:},
\ee
where $e^{:h_i:}$'s are symplectic polynomial maps and the numeral appearing in
the subscript indexes the polynomial maps. The polynomial maps are
determined by requiring that ${\calP}$ agree with ${\calM}$ up to
order 4. That is, when the $N$ polynomial maps are combined, the resulting
symplectic map should have all the monomials present in $f_3$ and $f_4$
with the correct coefficients up to order 4.

The basic idea in obtaining the required refactorization
is to group the monomial terms present in $f_3$ and $f_4$ [cf. Eq. (\ref{fs})]
such that the Lie transformation corresponding to each grouping gives a polynomial map.
It is obviously easy to handle monomials where both members of the
canonically conjugate pair are not present simultaneously. They can be grouped into
different polynomials (for example, monomials involving only the coordinate variables $q_i$'s
in one group and those involving only the momentum variables $p_i$'s in another group etc.) so that each one
of these is a polynomial of the first type. Since a product of two Lie transformations
($e^{:f_3:}$ and $e^{:f_4:}$) is being refactorized as a product of many
simpler polynomial symplectic maps, the coefficients multiplying the
monomials in each individual polynomial map will be in general different from
the coefficient multiplying the corresponding monomial in Eq. (\ref{fs}).
The relation between these coefficients is easily obtained using the
CBH theorem \cite{Cornwell}.
Monomials $s(z)$ where both
members of the canonically conjugate pair are present simultaneously
(like for example, $s(z)=q_1^3 p_1$) are more difficult to handle since
the corresponding Lie transformations $e^{:s(z):}$ typically give rise
to poles and branch points which we wish to avoid. But by
using a product of two polynomial maps of type 2 with carefully chosen
coefficients, these can also be generated.

Using the above procedure, it turns out that we require 23 polynomial maps for refactorization:
\be \label{polymap}
{\calM} \approx {\calP} = {\hat M} e^{:h_1:}\,e^{:h_2:} \cdots e^{:h_{23}:},
\ee
The $h_i$'s are given as follows:
\begin{eqnarray*} %\label{polymap3d}
h_1 & = & q_1^3\,b_{28} + q_1^2\,q_2\,b_{30} + q_1^2\,q_3\,b_{32} + q_1\,q_2^2\,b_{39} +
   q_1\,q_2\,q_3\,b_{41} + q_1\,q_3^2\,b_{46} + \\
& & q_2^3\,b_{64} + q_2^2\,q_3\,b_{66} +
  q_2\,q_3^2\,b_{71} + q_3^3\,b_{80} + q_1^4\,b_{84} + q_1^3\,q_2\,b_{86} + q_1^3\,q_3\,b_{88} +\\
& &
   q_1^2\,q_2^2\,b_{95} + q_1^2\,q_2\,q_3\,b_{97} + q_1^2\,q_3^2\,b_{102} +
 q_1\,q_2^3\,b_{120} +
   q_1\,q_2^2\,q_3\,b_{122} + \\
&&   q_1\,q_2\,q_3^2\,b_{127} + q_1\,q_3^3\,b_{136} + q_2^4\,b_{175} +
   q_2^3\,q_3\,b_{177} +
   q_2^2\,q_3^2\,b_{182} + \\
&& q_2\,q_3^3\,b_{191} + q_3^4\,b_{205}, \\
h_2 & = &
   \left[ (b_{29} + b_{34}) + q_2\,\left( b_{91} + b_{106} \right)  +
     p_2\,\left( b_{92} + b_{107} \right)  +
     q_3\,\left( b_{93} + b_{108} \right)  + \right. \\
&&     \left. p_3\,\left( b_{94} + b_{109} \right)  \right]
  {\left( p_1 + q_1 \right) }^3, \\
h_3 & = &
   \left[ (-b_{29} + b_{34}) + q_2\,\left( -b_{91} + b_{106} \right)  +
     p_2\,\left( -b_{92} + b_{107} \right)  + \right. \\
&& \left.     q_3\,\left( -b_{93} + b_{108} \right)  +
    p_3\,\left( -b_{94} + b_{109} \right)  \right]
 {\left( -p_1 + q_1 \right) }^3, \\
h_4 & = &
   \left[ (b_{65} + b_{68}) + q_1\,\left( b_{121} + b_{124} \right)  +
     p_1\,\left( b_{156} + b_{159} \right)  + \right. \\
&&  \left. q_3\,\left( b_{180} + b_{186} \right)  +
     p_3\,\left( b_{181} + b_{187} \right)  \right]
  {\left( p_2 + q_2 \right) }^3, \\
h_5 & = &
   \left[ (-b_{65} + b_{68}) + q_1\,\left( -b_{121} + b_{124} \right)  +
     p_1\,\left( -b_{156} + b_{159} \right)  + \right. \\
&& \left. q_3\,\left( -b_{180} + b_{186} \right)  +
 p_3\,\left( -b_{181} + b_{187} \right)  \right]
   {\left( -p_2 + q_2 \right) }^3, \\
h_6 & = &
   \left[ (b_{81} + b_{82}) + q_1\,\left( b_{137} + b_{138} \right)  +
     p_1\,\left( b_{172} + b_{173} \right)  +
     q_2\,\left( b_{192} + b_{193} \right)  +\right. \\
&&  \left.   p_2\,\left( b_{202} + b_{203} \right)  \right]
   {\left( p_3 + q_3 \right) }^3, \\
h_7 & = &
   \left[ (-b_{81} + b_{82}) + q_1\,\left( -b_{137} + b_{138} \right)  +
     p_1\,\left( -b_{172} + b_{173} \right)  +\right. \\
&&  \left.    q_2\,\left( -b_{192} + b_{193} \right)  +
   p_2\,\left( -b_{202} + b_{203} \right)  \right]
   {\left( -p_3 + q_3 \right) }^3, \\
h_8 & = & {\left( p_1 + q_1 \right) }^2\,
   \left( q_2\,b_{35} + q_3\,b_{37} + q_2^2\,b_{110} + q_2\,q_3\,b_{112} +
     p_3\,q_2\,b_{113} + q_3^2\,b_{117} \right), \\
h_9 & = & {\left( p_1 + q_1 \right) }^2\,
   \left( p_2\,b_{36} + p_3\,b_{38} +
     p_2^2\,b_{114} + p_2\,q_3\,b_{115} +
     p_2\,p_3\,b_{116} + p_3^2\,b_{119} \right), \\
h_{10} & = & {\left( p_2 + q_2 \right) }^2\,
   \left( q_1\,b_{40} + q_3\,b_{69} + q_1^2\,b_{96} + q_1\,q_3\,b_{125} +
     p_3\,q_1\,b_{126} + q_3^2\,b_{188} \right), \\
h_{11} & = & {\left( p_2 + q_2 \right) }^2\,
   \left( p_1\,b_{55} + p_3\,b_{70} +
     p_1^2\,b_{146} + p_1\,q_3\,b_{160} +
     p_1\,p_3\,b_{161} + p_3^2\,b_{190} \right), \\
h_{12} & = & {\left( p_3 + q_3 \right) }^2\,
   \left( q_1\,b_{47} + q_2\,b_{72} + q_1^2\,b_{103} + q_1\,q_2\,b_{128} +
     p_2\,q_1\,b_{134} + q_2^2\,b_{183} \right), \\
h_{13} & = & {\left( p_3 + q_3 \right) }^2\,
   \left( p_1\,b_{62} + p_2\,b_{78} +
     p_1^2\,b_{153} + p_1\,q_2\,b_{163} +
     p_1\,p_2\,b_{169} + p_2^2\,b_{199} \right), \\
h_{14} & = & p_2\,q_1^2\,b_{31} + p_3\,q_1^2\,b_{33} +
   p_2^2\,q_1\,b_{43} + p_2\,p_3\,q_1\,b_{45} +
   p_3^2\,q_1\,b_{48} + p_2\,q_1^3\,b_{87} + \\
&&   p_3\,q_1^3\,b_{89} +
   p_2^2\,q_1^2\,b_{99} +
   p_2\,p_3\,q_1^2\,b_{101} +
   p_3^2\,q_1^2\,b_{104} + p_2^3\,q_1\,b_{130} + \\
&&   p_2^2\,p_3\,q_1\,b_{132} +
   p_2\,p_3^2\,q_1\,b_{135} + p_3^3\,q_1\,b_{139}, \\
h_{15} & = & p_1^2\,q_2\,b_{50} + p_1^2\,q_3\,b_{52} +
   p_1\,q_2^2\,b_{54} + p_1\,q_2\,q_3\,b_{56} +
   p_1\,q_3^2\,b_{61} + p_1^3\,q_2\,b_{141} + \\
&&   p_1^3\,q_3\,b_{143} +
  p_1^2\,q_2^2\,b_{145} +
   p_1^2\,q_2\,q_3\,b_{147} + p_1^2\,q_3^2\,b_{152} +
   p_1\,q_2^3\,b_{155} + \\
&&  p_1\,q_2^2\,q_3\,b_{157} +
   p_1\,q_2\,q_3^2\,b_{162} + p_1\,q_3^3\,b_{171}, \\
h_{16} & = & p_1\,p_3\,q_2\,b_{57} + p_3\,q_2^2\,b_{67} +
   p_3^2\,q_2\,b_{73} +
   p_1^2\,p_3\,q_2\,b_{148} +
   p_1\,p_3\,q_2^2\,b_{158} + \\
&&   p_1\,p_3^2\,q_2\,b_{164} +
   p_3\,q_2^3\,b_{178} + p_3^2\,q_2^2\,b_{184} +
   p_3^3\,q_2\,b_{194}, \\
h_{17} & = & p_2\,q_1\,q_3\,b_{44} + p_2^2\,q_3\,b_{75} +
   p_2\,q_3^2\,b_{77} + p_2\,q_1^2\,q_3\,b_{100} +
   p_2^2\,q_1\,q_3\,b_{131} +  \\
& &   p_2\,q_1\,q_3^2\,b_{133} +
   p_2^3\,q_3\,b_{196} + p_2^2\,q_3^2\,b_{198} +
   p_2\,q_3^3\,b_{201}, \\
h_{18} & = & p_1\,p_2\,q_3\,b_{59} +
   p_1^2\,p_2\,q_3\,b_{150} +
   p_1\,p_2^2\,q_3\,b_{166} +
   p_1\,p_2\,q_3^2\,b_{168}, \\
h_{19} & = & p_3\,q_1\,q_2\,b_{42} + p_3\,q_1^2\,q_2\,b_{98} +
   p_3\,q_1\,q_2^2\,b_{123} + p_3^2\,q_1\,q_2\,b_{129}, \\
h_{20} & = & p_1^3\,b_{49} + p_1^2\,p_2\,b_{51} +
   p_1^2\,p_3\,b_{53} +
   p_1\,p_2^2\,b_{58} +
   p_1\,p_2\,p_3\,b_{60} +
   p_1\,p_3^2\,b_{63} + \\
&&   p_2^3\,b_{74} +
   p_2^2\,p_3\,b_{76} +
   p_2\,p_3^2\,b_{79} + p_3^3\,b_{83} +
   p_1^4\,b_{140} + p_1^3\,p_2\,b_{142} + \\
&&   p_1^3\,p_3\,b_{144} +
   p_1^2\,p_2^2\,b_{149} +
   p_1^2\,p_2\,p_3\,b_{151} +
   p_1^2\,p_3^2\,b_{154} +
   p_1\,p_2^3\,b_{165} + \\
&&   p_1\,p_2^2\,p_3\,b_{167} +
   p_1\,p_2\,p_3^2\,b_{170} +
   p_1\,p_3^3\,b_{174}+
   p_2^4\,b_{195} +
   p_2^3\,p_3\,b_{197} + \\
&&   p_2^2\,p_3^2\,b_{200} +
   p_2\,p_3^3\,b_{204} + p_3^4\,b_{209}, \\
h_{21} & = & {\left( p_1 + q_1 + p_1^2\,b_{105} \right) }^3 +
   {\left( p_2 + q_2 + p_2^2\,b_{185} \right) }^3 +
   {\left( p_3 + q_3 + p_3^2\,b_{208} \right) }^3, \\
h_{22} & = & {\left( -p_1 - q_1 + q_1^2\,b_{85} \right) }^3 +
   {\left( -p_2 - q_2 + q_2^2\,b_{176} \right) }^3 + \\
&&   {\left( -p_3 - q_3 + q_3^2\,b_{206} \right) }^3, \\
h_{23} & = & {\left( p_1 + q_1 \right) }^4\,b_{90} +
   {\left( p_1 + q_1 \right) }^2\,
    {\left( p_2 + q_2 \right) }^2\,b_{111} +
   {\left( p_1 + q_1 \right) }^2\,
    {\left( p_3 + q_3 \right) }^2\,b_{118} \\
&&   +{\left( p_2 + q_2 \right) }^4\,b_{179} +
   {\left( p_2 + q_2 \right) }^2\,
    {\left( p_3 + q_3 \right) }^2\,b_{189} +
   {\left( p_3 + q_3 \right) }^4\,b_{207}.
\end{eqnarray*}
Here $b_i$'s are at present unknown coefficients. As mentioned above,
by forcing the refactorized form ${\calP}$ to equal the original map ${\calM}$
up to order 4 and using the CBH
theorem\cite{Cornwell}, we can easily compute these unknown coefficients in terms of
the known $a_i$'s. These expressions are available from the author as part of a FORTRAN
program implementing the above algorithm.

The explicit actions of the polynomial maps on phase space variables
can be obtained and they are given below. This completely
determines the refactorized map ${\calP}$. Each $\exp(:h_i:)$ is a
polynomial map which can be evaluated exactly and is explicitly symplectic.
Thus by using ${\calP}$ instead of ${\calM}$ in Eq. (\ref{rec}), we obtain
an explicitly symplectic integration algorithm. Further, it is fast to
evaluate and does not introduce spurious poles and branch points.
The above factorization is not unique. However, the principles outlined
earlier impose restrictions on the possible forms and this eases
considerably the task of refactorization. Moreover, we require the
coefficients $b_i$ to be polynomials in the known coefficients $a_i$.
Otherwise this can lead to divergences when $a_i$'s take on certain
special values. Finally, we minimize the number of polynomial maps
in the refactorized form. Our studies show that different polynomial
map refactorizations obeying the above restrictions do not lead to
any significant differences in their behavior.

We now derive the explicit actions of the polynomial maps
on phase space variables. First consider $\exp(:h_1:)$. We obtain
its action on the phase space variables by following the procedure
outlined in the paragraph before Eq. (\ref{Heqns}). We notice
that $h_1 = h_1(q_1,q_2,q_3)$ depends only on the coordinate
variables which are therefore cyclic variables. Hence we
immediately obtain: $q_i(t) = q_i(0)$ ($i=1,2,3$). Solving
the Hamilton's equations of motion for $p_i$ with $h_1$ as the
Hamiltonian from $t=0$ to $t=-1$ we get:
\be
p_i(-1) = p_i(0)+ \frac{\partial h_1}{\partial q_i}(q_1(0),q_2(0),q_3(0)),
\ \ \ i=1,2,3.
\ee
Denoting $z_i(0)$ and $z_i(-1)$ by $z_i^{in}$ and
$z_i^{fin}$ respectively, we finally obtain the action of $\exp(:h_1:)$:
\be
q_i^{fin} = q_i^{in}; \ \ \ p_i^{fin} = p_i^{in}+
\frac{\partial h_1}{\partial q_i}(q_1^{in},q_2^{in},q_3^{in}),
\ \ \ i=1,2,3.
\ee

Next consider the action of $\exp(:h_2:)$. The 2 factors in the product
are independently conserved under the Hamiltonian flow. Consequently,
we have
\beq
A_2 & = & \left[ (b_{29} + b_{34}) + q_2\,\left( b_{91} + b_{106} \right)  +
     p_2\,\left( b_{92} + b_{107} \right)  + \right. \nonumber \\
&&   \left. q_3\,\left( b_{93} + b_{108} \right)  +
     p_3\,\left( b_{94} + b_{109} \right)  \right] \nonumber \\
& = & \left[ (b_{29} + b_{34}) + q_2(0)\,\left( b_{91} + b_{106} \right)  +
     p_2(0)\,\left( b_{92} + b_{107} \right)  +\right. \nonumber \\
&& \left.     q_3(0)\,\left( b_{93} + b_{108} \right)  +
   p_3(0)\,\left( b_{94} + b_{109} \right)  \right], \\
B_2 & = & (q_1+p_1) = (q_1(0)+p_1(0)). \nonumber
\eeq
Solving Hamilton's equation of motion with $h_2$ as the Hamiltonian
from $t=0$ to $t=-1$ we get
\beq
q_1(-1) & = & q_1(0)-3A_2 B_2^2; \ \ \ p_1(-1)  =  p_1(0)+3A_2 B_2^2; \nonumber \\
q_2(-1) & = & q_2(0)-B_2^3 \left( b_{92} + b_{107} \right); \ \ \
p_2(-1)  =  p_2(0)+B_2^3 \left( b_{91} + b_{106} \right); \\
q_3(-1) & = & q_3(0)-B_2^3 \left( b_{94} + b_{109} \right); \ \ \
p_3(-1)  =  p_3(0)+B_2^3 \left( b_{93} + b_{108} \right). \nonumber
\eeq
Denoting $z_i(0)$ and $z_i(-1)$ by $z_i^{in}$ and
$z_i^{fin}$ respectively, we obtain the action of $\exp(:h_2:)$:
\beq
q_1^{fin} & = & q_1^{in}-3A_2 B_2^2; \ \ \
p_1^{fin}  =  p_1^{in}+3A_2 B_2^2; \nonumber \\
q_2^{fin} & = & q_2^{in}-B_2^3 \left( b_{92} + b_{107} \right); \ \ \
p_2^{fin}  =  p_2^{in}+B_2^3 \left( b_{91} + b_{106} \right); \\
q_3^{fin} & = & q_3^{in}-B_2^3 \left( b_{94} + b_{109} \right); \ \ \
p_3^{fin}  =  p_3^{in}+B_2^3 \left( b_{93} + b_{108} \right), \nonumber
\eeq
where $A_2,B_2$ are now functions of $z^{in}$.
The actions of $\exp(:h_i:)$, $i=3,4, \ldots ,7$ on the phase space
variables are obtained in a similar fashion and these actions are
listed in the Appendix.

We now consider the action of $\exp(:h_8:)$.
From the Hamilton's equations of motion, we have the following
conserved quantities:
\beq
A_8 & = & \left( q_2\,b_{35} + q_3\,b_{37} + q_2^2\,b_{110} + q_2\,q_3\,b_{112} +
     p_3\,q_2\,b_{113} + q_3^2\,b_{117} \right) \nonumber \\
& = & \left( q_2(0)b_{35} + q_3(0)b_{37} + q_2^2(0)b_{110} + q_2(0)q_3(0)b_{112} +
      \right. \nonumber \\
&& \left. p_3(0)q_2(0)b_{113} +q_3^2(0)b_{117} \right), \\
B_8 & = & (q_1+p_1) = (q_1(0)+p_1(0)). \nonumber
\eeq
Solving the equations of motion for $q_i$'s with $h_8$ as the Hamiltonian from
$t=0$ to $t$ we get:
\beq
q_1(t) & = & q_1(0)+2 A_8 B_8 t; \ \ \ q_2(t) = q_2(0); \nonumber \\
q_3(t) & = & q_3(0)
+ B_8^2 b_{113} q_2(0)t.
\eeq
For the momentum variables we get the following differential equations:
\beq
\frac{d p_1(t)}{dt} & = & -2 A_8 B_8, \nonumber \\
\frac{d p_2(t)}{dt} & = & - B_8^2 [b_{35}+2 b_{110} q_2(t) +
b_{112} q_3(t) + b_{113} p_3(t)],  \\
\frac{d p_3(t)}{dt} & = & - B_8^2 [b_{37}+ b_{112} q_2(t) + 2 b_{117} q_3(t)].  \nonumber
\eeq
The first equation can be trivially solved to obtain $p_1(t)$. After substituting
for $q_2(t)$ and $q_3(t)$ which are known, we can next solve the last equation for
$p_3(t)$. Substituting this in the second equation, we finally get $p_2(t)$.
Setting $t=-1$ and denoting $z_i(0)$, $z_i(-1)$ by $z_i^{in}$,
$z_i^{fin}$ respectively, we obtain the action of $\exp(:h_8:)$:
\beq
q_1^{fin} & = & q_1^{in}-2 A_8 B_8; \ \ \
p_1^{fin}  =  p_1^{in}+ 2 A_8 B_8; \nonumber \\
q_2^{fin} & = & q_2^{in} ; \ \ \
p_2^{fin}  =  p_2^{in}+B_8^2 \left[ b_{35}+2 b_{110} q_2^{in} +
b_{112} q_3^{in} + b_{113} p_3^{in} + \right. \nonumber \\
&& \left. B_8^2 b_{113} (b_{37}/2+b_{113} q_3^{in} - B_8^2 b_{113} b_{117} q_2^{in}/3) \right]; \\
q_3^{fin} & = & q_3^{in}-B_8^2 b_{113} q_2^{in}; \nonumber \\
p_3^{fin} & = & p_3^{in}+B_8^2 \left[ b_{37}+ b_{112} q_2^{in}
+ 2 b_{117} q_3^{in} - B_8^2 b_{113} b_{117} q_2^{in} \right], \nonumber
\eeq
where $A_8,B_8$ are now functions of $z^{in}$.
The actions of $\exp(:h_i:)$, $i=9,10, \ldots $, $13$ on the phase space
variables are obtained in a similar fashion and these actions are
listed in the Appendix.

Next consider the action of $\exp(:h_{14}:)$. We notice
that $h_{14} = h_{14}(q_1,p_2,p_3)$ is independent of $p_1$,
$q_2$ and $q_3$. Hence $q_1$, $p_2$, $p_3$ are cyclic variables
and are conserved under the action of the Hamiltonian. Solving
the Hamilton's equations of motion for $p_1$,
$q_2$ and $q_3$ with $h_{14}$ as the
Hamiltonian from $t=0$ to $t=-1$ we get:
\beq
p_1(-1) & = & p_1(0)+ \frac{\partial h_{14}}{\partial q_1}(q_1(0),p_2(0),p_3(0)), \nonumber \\
q_2(-1) & = & q_2(0)- \frac{\partial h_{14}}{\partial p_2}(q_1(0),p_2(0),p_3(0)), \\
q_3(-1) & = & q_3(0)- \frac{\partial h_{14}}{\partial p_3}(q_1(0),p_2(0),p_3(0)). \nonumber
\eeq
Denoting $z_i(0)$ and $z_i(-1)$ by $z_i^{in}$ and
$z_i^{fin}$ respectively, we finally obtain the action of $\exp(:h_{14}:)$:
\beq
q_1^{fin} & = & q_1^{in}; \ \ \ p_1^{fin} = p_1^{in}+
\frac{\partial h_{14}}{\partial q_1}(q_1^{in},p_2^{in},p_3^{in}), \nonumber \\
q_2^{fin} & = & q_2^{in}-
\frac{\partial h_{14}}{\partial p_2}(q_1^{in},p_2^{in},p_3^{in}), \ \ \
p_2^{fin} = p_2^{in}, \\
q_3^{fin} & = & q_3^{in}-
\frac{\partial h_{14}}{\partial p_3}(q_1^{in},p_2^{in},p_3^{in}), \ \ \
p_3^{fin} = p_3^{in}. \nonumber
\eeq
The actions of $\exp(:h_i:)$, $i=15,16, \ldots ,20$ on the phase space
variables are obtained in a similar fashion and these actions are
listed in the Appendix.

We now consider the action of $\exp(:h_{21}:)$.
From the Hamilton's equations of motion, we have the following
conserved quantities:
\beq
A_{21} & = & {\left( p_1 + q_1 + p_1^2\,b_{105} \right) }
= {\left( p_1(0) + q_1(0) + p_1^2(0)\,b_{105} \right) }, \nonumber \\
B_{21} & = & {\left( p_2 + q_2 + p_2^2\,b_{185} \right) } =
{\left( p_2(0) + q_2(0) + p_2^2(0)\,b_{185} \right) } \\
C_{21} & = & {\left( p_3 + q_3 + p_3^2\,b_{208} \right) } =
{\left( p_3(0) + q_3(0) + p_3^2(0)\,b_{208} \right) }. \nonumber
\eeq
Solving the equations of motion for $p_i$'s with $h_{21}$ as the Hamiltonian from
$t=0$ to $t$ we get:
\beq
p_1(t)&=&p_1(0)-3 A_{21}^2 t ; \ \ \ p_2(t)=p_2(0)-3 B_{21}^2 t ; \nonumber \\
p_3(t)&=&p_3(0)-3 C_{21}^2 t.
\eeq
The equations of motion for $q_i$'s are given as:
\beq
\frac{d q_1(t)}{dt} & = & 3 A_{21}^2 (1+2 b_{105} p_1(t)), \nonumber \\
\frac{d q_2(t)}{dt} & = & 3 B_{21}^2 (1+2 b_{185} p_2(t)),  \\
\frac{d q_3(t)}{dt} & = & 3 B_{21}^2 (1+2 b_{208} p_3(t)). \nonumber
\eeq
Substituting the expressions for $p_1(t)$, $p_2(t)$, $p_3(t)$
obtained earlier, the above equations can be easily solved.
Setting $t=-1$ and denoting $z_i(0)$, $z_i(-1)$ by $z_i^{in}$,
$z_i^{fin}$ respectively, we obtain the action of $\exp(:h_{21}:)$:
\beq
q_1^{fin} & = & q_1^{in}-3 A_{21}^2 (1+2 b_{105} p_1^{in}+ 3 A_{21}^2 b_{105}); \ \ \
p_1^{fin}  =  p_1^{in}+ 3 A_{21}^2; \nonumber \\
q_2^{fin} & = & q_2^{in}-3 B_{21}^2 (1+2 b_{185} p_2^{in}+ 3 B_{21}^2 b_{185}); \ \ \
p_2^{fin}  =  p_2^{in}+ 3 B_{21}^2; \nonumber \\
q_3^{fin} & = & q_3^{in}-3 C_{21}^2 (1+2 b_{208} p_3^{in}+ 3 C_{21}^2 b_{208}); \ \ \
p_3^{fin}  =  p_3^{in}+ 3 C_{21}^2, \nonumber
\eeq
where $A_{21},B_{21},C_{21}$ are now functions of $z^{in}$.
The action of $\exp(:h_{22}:)$ on the phase space
variables is obtained in a similar fashion and is
listed in the Appendix.

Finally, we consider the action of $\exp(:h_{23}:)$.
From the Hamilton's equations of motion, we have the following
conserved quantities:
\beq
A_{23} & = & {\left( p_1 + q_1 \right) }
= {\left( p_1(0) + q_1(0)  \right) }, \nonumber \\
B_{23} & = & {\left( p_2 + q_2 \right) } =
{\left( p_2(0) + q_2(0)  \right) }, \\
C_{23} & = & {\left( p_3 + q_3 \right) } =
{\left( p_3(0) + q_3(0)  \right) }. \nonumber
\eeq
From the equations of motion with $h_{23}$ as the Hamiltonian we
get:
\beq
\frac{d q_1(t)}{dt} & = & - \frac{d p_1(t)}{dt} =
4 b_{90} A_{23}^3 + 2 b_{111} A_{23} B_{23}^2 + 2 b_{118} A_{23} C_{23}^2, \nonumber \\
\frac{d q_2(t)}{dt} & = & - \frac{d p_2(t)}{dt} =
4 b_{179} B_{23}^3 + 2 b_{111} A_{23}^2 B_{23} + 2 b_{189} B_{23} C_{23}^2, \nonumber \\
\frac{d q_3(t)}{dt} & = & - \frac{d p_3(t)}{dt} =
4 b_{207} C_{23}^3 + 2 b_{118} A_{23}^2 C_{23} + 2 b_{189} B_{23}^2 C_{23}.
\eeq
Solving these equations from $t=0$ to $t=-1$ and denoting $z_i(0)$, $z_i(-1)$ by $z_i^{in}$,
$z_i^{fin}$ respectively, we obtain the action of $\exp(:h_{23}:)$:
\beq
q_1^{fin} & = & q_1^{in}-[4 b_{90} A_{23}^3 + 2 b_{111} A_{23} B_{23}^2 + 2 b_{118} A_{23} C_{23}^2]; \nonumber \\
p_1^{fin} & = & p_1^{in}+ [4 b_{90} A_{23}^3 + 2 b_{111} A_{23} B_{23}^2 + 2 b_{118} A_{23} C_{23}^2]; \nonumber \\
q_2^{fin} & = & q_2^{in}-[4 b_{179} B_{23}^3 + 2 b_{111} A_{23}^2 B_{23} + 2 b_{189} B_{23} C_{23}^2]; \nonumber \\
p_2^{fin} & = & p_2^{in}+ [4 b_{179} B_{23}^3 + 2 b_{111} A_{23}^2 B_{23} + 2 b_{189} B_{23} C_{23}^2];  \\
q_3^{fin} & = & q_3^{in}-[4 b_{207} C_{23}^3 + 2 b_{118} A_{23}^2 C_{23} + 2 b_{189} B_{23}^2 C_{23}]; \nonumber \\
p_3^{fin} & = & p_3^{in}+ [4 b_{207} C_{23}^3 + 2 b_{118} A_{23}^2 C_{23} + 2 b_{189} B_{23}^2 C_{23}], \nonumber
\eeq
where $A_{23},B_{23},C_{23}$ are now functions of $z^{in}$.

Substituting in Eq. (\ref{polymap}) the explicit formulas for the actions of the polynomial maps
listed above and in the Appendix, we can evaluate the action of ${\calP}$ without violating the
symplectic condition. Using this explicitly symplectic map in Eq. (\ref{rec}),
we have the desired symplectic integration algorithm.

\section{Applications}
\noindent
We have applied the method to a large
particle storage ring for storing charged particles. This
storage ring consists of 5109 individual elements (where these elements
could be drifts, bending
magnets, quadrupoles or sextupoles). If one tries to numerically
integrate the trajectory of a charged particle through this ring
using a conventional integration algorithm,
one has to go through the ring element by element where each element
is described by its own Hamiltonian. This is cumbersome and slow
and further, does not respect the Hamiltonian nature of the system.
On the other hand, a map based approach where
one represents the entire storage ring in terms of a single map is much faster \cite{yan,berg}.
When this is combined with our polynomial map refactorization, one obtains a
symplectic integration algorithm which is both fast and accurate and is
ideally suited for such complex real life systems. The $q_1 - p_1$
phase plot for one million turns around the ring using our polynomial map method
is given in Figure 1. In this case, $q_1$ and $p_1$ represent the deviations
from the closed orbit coordinate and momentum respectively. From theoretical
considerations, we expect the so-called betatron oscillations in these
variables. This manifests itself as ellipses in the phase space plot
of $q_1$ and $p_1$ variables. In Figure 1, we observe the expected
betatron oscillations. We also see the thickening of the ellipses
caused by nonlinearities present in the sextupoles.

\begin{figure}[htbp] %ORIGINAL SIZE: width=1.4TRUEIN; height=1.5TRUEIN
\vspace*{13pt}
\centerline{\psfig{file=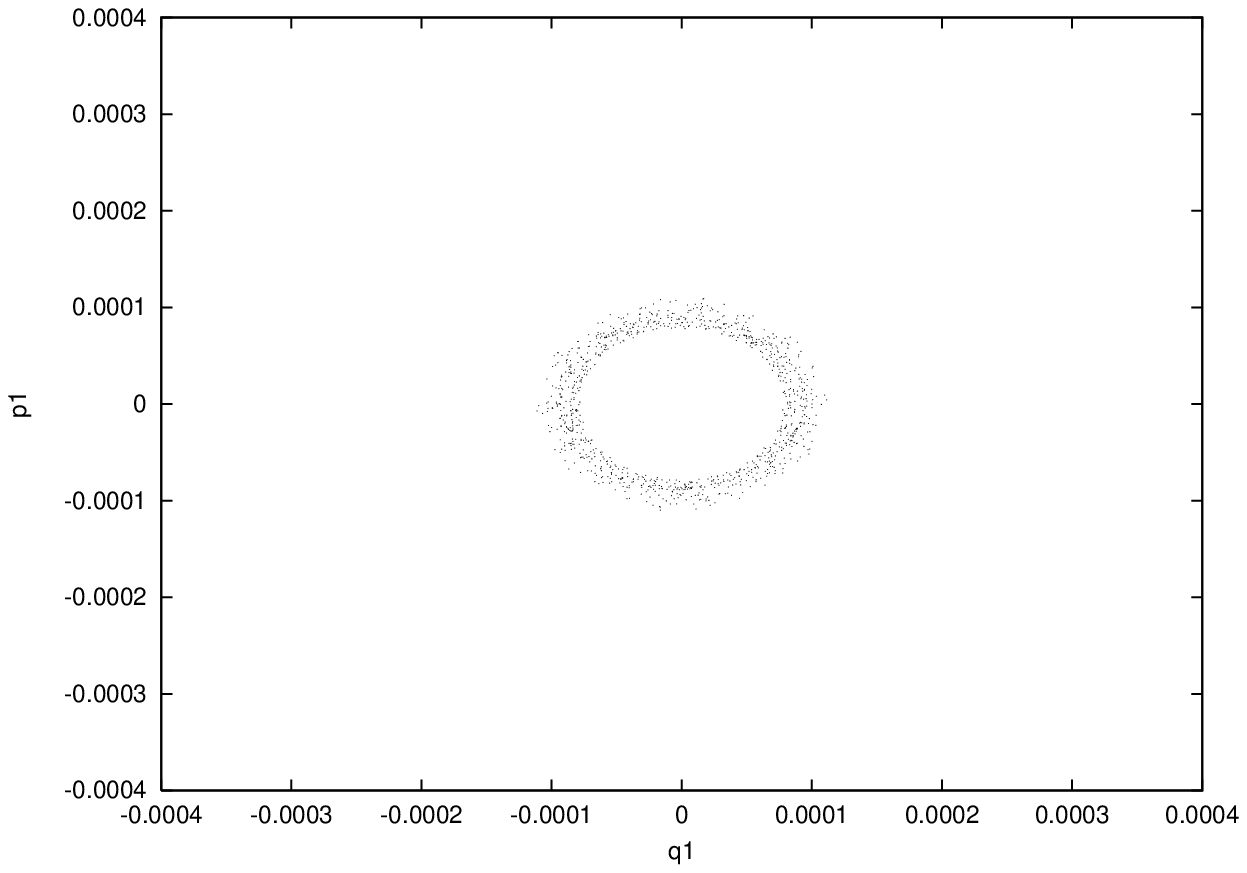}} %100 percent
\vspace*{13pt}
\fcaption{This figure shows the $q_1 - p_1$ phase
space plot for one million turns around a storage ring using
the polynomial map method (only every 1000th point is plotted).}
\end{figure}

\section{Conclusions}
\noindent
To conclude, we described in detail a new symplectic integration algorithm based
on polynomial map refactorization. We enumerated the types of symplectic
maps which give rise to polynomial actions on phase space variables.
For a six dimensional phase space, we obtained the refactorization of
a given symplectic map in terms of 23 polynomial maps. The explicit
actions of these polynomial maps were derived.
This polynomial map method can be used to study long
term stability of complicated nonlinear Hamiltonian
systems.

\nonumsection{Acknowledgements}
\noindent
This work was supported by the Homi Bhabha Fellowship. The author is
also associated with the Jawaharlal Nehru Center for Advanced
Scientific Research as a honorary faculty member.

\appendix

\noindent
The actions of polynomial maps $\exp(:h_i:)$ which were not
listed in the main text are given in this Appendix.

The action of $\exp(:h_3:)$ is given as follows:
\beq
q_1^{fin} & = & q_1^{in}+3A_3 B_3^2; \ \ \
p_1^{fin}  =  p_1^{in}+3A_3 B_3^2; \nonumber \\
q_2^{fin} & = & q_2^{in}-B_3^3 \left( -b_{92} + b_{107} \right); \ \ \
p_2^{fin}  =  p_2^{in}+B_3^3 \left( -b_{91} + b_{106} \right); \\
q_3^{fin} & = & q_3^{in}-B_3^3 \left( -b_{94} + b_{109} \right); \ \ \
p_3^{fin}  =  p_3^{in}+B_3^3 \left( -b_{93} + b_{108} \right), \nonumber
\eeq
where
\beq
A_3 & = & \left[ (-b_{29} + b_{34}) + q_2^{in}\,\left( -b_{91} + b_{106} \right)  +
     p_2^{in}\,\left( -b_{92} + b_{107} \right)  + \right. \nonumber \\
&&  \left.     q_3^{in}\,\left( -b_{93} + b_{108} \right)  +
 p_3^{in}\,\left( -b_{94} + b_{109} \right)  \right], \\
B_3 & = &  (q_1^{in}-p_1^{in}). \nonumber
\eeq

The action of $\exp(:h_4:)$ is given as follows:
\beq
q_1^{fin} & = & q_1^{in}-B_4^3 \left( b_{156} + b_{159} \right); \ \ \
p_1^{fin}  =  p_1^{in}+B_4^3 \left( b_{121} + b_{124} \right); \nonumber \\
q_2^{fin} & = & q_2^{in}-3A_4 B_4^2; \ \ \
p_2^{fin}  =  p_2^{in}+3A_4 B_4^2; \\
q_3^{fin} & = & q_3^{in}-B_4^3 \left( b_{181} + b_{187} \right); \ \ \
p_3^{fin}  =  p_3^{in}+B_4^3 \left( b_{180} + b_{186} \right), \nonumber
\eeq
where
\beq
A_4 & = & \left[ (b_{65} + b_{68}) + q_1^{in}\,\left( b_{121} + b_{124} \right)  +
     p_1^{in}\,\left( b_{156} + b_{159} \right)  +\right. \nonumber \\
&&  \left.     q_3^{in}\,\left( b_{180} + b_{186} \right)  +
 p_3^{in}\,\left( b_{181} + b_{187} \right)  \right], \\
B_4 & = &  (q_2^{in}+p_2^{in}). \nonumber
\eeq

The action of $\exp(:h_5:)$ is given as follows:
\beq
q_1^{fin} & = & q_1^{in}-B_5^3 \left( -b_{156} + b_{159} \right); \ \ \
p_1^{fin}  =  p_1^{in}+B_5^3 \left( -b_{121} + b_{124} \right); \nonumber \\
q_2^{fin} & = & q_2^{in}+3A_5 B_5^2; \ \ \
p_2^{fin}  =  p_2^{in}+3A_5 B_5^2; \\
q_3^{fin} & = & q_3^{in}-B_5^3 \left( -b_{181} + b_{187} \right); \ \ \
p_3^{fin}  =  p_3^{in}+B_5^3 \left( -b_{180} + b_{186} \right), \nonumber
\eeq
where
\beq
A_5 & = & \left[ (-b_{65} + b_{68}) + q_1^{in}\,\left( -b_{121} + b_{124} \right)  +
     p_1^{in}\,\left( -b_{156} + b_{159} \right)  +\right. \nonumber \\
&&  \left.     q_3^{in}\,\left( -b_{180} + b_{186} \right)  +
 p_3^{in}\,\left( -b_{181} + b_{187} \right)  \right], \\
B_5 & = &  (q_2^{in}-p_2^{in}). \nonumber
\eeq

The action of $\exp(:h_6:)$ is given as follows:
\beq
q_1^{fin} & = & q_1^{in}-B_6^3 \left( b_{172} + b_{173} \right); \ \ \
p_1^{fin}  =  p_1^{in}+B_6^3 \left( b_{137} + b_{138} \right); \nonumber \\
q_2^{fin} & = & q_2^{in}-B_6^3 \left( b_{202} + b_{203} \right); \ \ \
p_2^{fin}  =  p_2^{in}+B_6^3 \left( b_{192} + b_{193} \right); \\
q_3^{fin} & = & q_3^{in}-3A_6 B_6^2; \ \ \
p_3^{fin}  =  p_3^{in}+3A_6 B_6^2, \nonumber
\eeq
where
\beq
A_6 & = & \left[ (b_{81} + b_{82}) + q_1^{in}\,\left( b_{137} + b_{138} \right)  +
     p_1^{in}\,\left( b_{172} + b_{173} \right)  +\right. \nonumber \\
&&  \left.     q_2^{in}\,\left( b_{192} + b_{193} \right)  +
 p_2^{in}\,\left( b_{202} + b_{203} \right)  \right], \\
B_6 & = &  (q_3^{in}+p_3^{in}). \nonumber
\eeq

The action of $\exp(:h_7:)$ is given as follows:
\beq
q_1^{fin} & = & q_1^{in}-B_7^3 \left( -b_{172} + b_{173} \right); \ \ \
p_1^{fin}  =  p_1^{in}+B_7^3 \left( -b_{137} + b_{138} \right); \nonumber \\
q_2^{fin} & = & q_2^{in}-B_7^3 \left( -b_{202} + b_{203} \right); \ \ \
p_2^{fin}  =  p_2^{in}+B_7^3 \left( -b_{192} + b_{193} \right); \\
q_3^{fin} & = & q_3^{in}+3A_7 B_7^2; \ \ \
p_3^{fin}  =  p_3^{in}+3A_7 B_7^2, \nonumber
\eeq
where
\beq
A_7 & = & \left[ (-b_{81} + b_{82}) + q_1^{in}\,\left( -b_{137} + b_{138} \right)  +
     p_1^{in}\,\left( -b_{172} + b_{173} \right)  +\right. \nonumber \\
&&  \left.     q_2^{in}\,\left( -b_{192} + b_{193} \right)  +
 p_2^{in}\,\left( -b_{202} + b_{203} \right)  \right], \\
B_7 & = &  (q_3^{in}+-p_3^{in}). \nonumber
\eeq

The action of $\exp(:h_9:)$ is given as follows:
\beq
q_1^{fin} & = & q_1^{in}-2 A_9 B_9; \ \ \
p_1^{fin}  =  p_1^{in}+ 2 A_9 B_9; \nonumber \\
q_2^{fin} & = & q_2^{in}+B_9^2 \left[ -b_{36}-2 b_{114} p_2^{in} -
b_{115} q_3^{in} - b_{116} p_3^{in} + \right. \nonumber \\
&& \left. B_9^2 b_{115} (b_{38}/2+b_{119} p_3^{in} + B_9^2 b_{115} b_{119} p_2^{in}/3) \right]; \ \ \
p_2^{fin}  =  p_2^{in} \\
q_3^{fin} & = & q_3^{in}-B_9^2 \left[ b_{38}+ b_{116} p_2^{in}
+ 2 b_{119} p_3^{in} + B_9^2 b_{115} b_{119} p_2^{in} \right]; \nonumber \\
p_3^{fin}  &= & p_3^{in}+B_9^2 b_{115} p_2^{in}, \nonumber
\eeq
where
\beq
A_9 & = & \left( p_2^{in}\,b_{36} + p_3^{in}\,b_{38} + (p_2^{in})^2\,b_{114}
+ p_2^{in}\,q_3^{in}\,b_{115} +
     p_2^{in}\,p_3^{in}\,b_{116} + \right. \nonumber \\
&& \left. (p_3^{in})^2\,b_{119} \right), \\
B_9 & = & (q_1^{in}+p_1^{in}). \nonumber
\eeq

The action of $\exp(:h_{10}:)$ is given as follows:
\beq
q_1^{fin} & = & q_1^{in}; \ \ \
p_1^{fin}  =  p_1^{in}+B_{10}^2 \left[ b_{40}+2 b_{96} q_1^{in} +
b_{125} q_3^{in} + b_{126} p_3^{in} + \right. \nonumber \\
&& \left. B_{10}^2 b_{126} (b_{69}/2+b_{188} q_3^{in}
- B_{10}^2 b_{126} b_{188} q_1^{in}/3) \right]; \nonumber \\
q_2^{fin} & = & q_2^{in}-2 A_{10} B_{10} ; \ \ \
p_2^{fin}  =  p_2^{in}+ 2 A_{10} B_{10}; \\
q_3^{fin} & = & q_3^{in}-B_{10}^2 b_{126} q_1^{in}; \nonumber \\
p_3^{fin}  &= & p_3^{in}+B_{10}^2 \left[ b_{69}+ b_{125} q_1^{in}
+ 2 b_{188} q_3^{in} - B_{10}^2 b_{126} b_{188} q_1^{in} \right], \nonumber
\eeq
where
\beq
A_{10} & = & \left( q_1^{in}\,b_{40} + q_3^{in}\,b_{69} + (q_1^{in})^2\,b_{96}
+ q_1^{in}\,q_3^{in}\,b_{125} +
     p_3^{in}\,q_1^{in}\,b_{126} \right. \nonumber \\
&& \left. + (q_3^{in})^2\,b_{188} \right) \\
B_{10} & = & (q_2^{in}+p_2^{in}). \nonumber
\eeq

The action of $\exp(:h_{11}:)$ is given as follows:
\beq
q_1^{fin} & = & q_1^{in}+B_{11}^2 \left[ -b_{55}-2 b_{146} p_1^{in} -
b_{160} q_3^{in} - b_{161} p_3^{in} + \right. \nonumber \\
&& \left. B_{11}^2 b_{160} (b_{70}/2+b_{190} p_3^{in}
+ B_{11}^2 b_{160} b_{190} p_1^{in}/3) \right]; \ \ \
p_1^{fin}  =  p_1^{in}; \nonumber \\
q_2^{fin} & = & q_2^{in}-2 A_{11} B_{11}; \ \ \
p_2^{fin}  =  p_2^{in}+ 2 A_{11} B_{11} \\
q_3^{fin} & = & q_3^{in}-B_{11}^2 \left[ b_{70}+ b_{161} p_1^{in}
+ 2 b_{190} p_3^{in} + B_{11}^2 b_{160} b_{190} p_1^{in} \right]; \nonumber \\
p_3^{fin} & = & p_3^{in}+B_{11}^2 b_{160} p_1^{in}, \nonumber
\eeq
where
\beq
A_{11} & = & \left( p_1^{in}\,b_{55} + p_3^{in}\,b_{70} + (p_1^{in})^2\,b_{146}
+ p_1^{in}\,q_3^{in}\,b_{160} +
     p_1^{in}\,p_3^{in}\,b_{161} + \right. \nonumber \\
&& \left. (p_3^{in})^2\,b_{190} \right), \\
B_{11} & = & (q_2^{in}+p_2^{in}). \nonumber
\eeq

The action of $\exp(:h_{12}:)$ is given as follows:
\beq
q_1^{fin} & = & q_1^{in}; \ \ \
p_1^{fin}  =  p_1^{in}+B_{12}^2 \left[ b_{47}+2 b_{103} q_1^{in} +
b_{128} q_2^{in} + b_{134} p_2^{in} + \right. \nonumber \\
&& \left. B_{12}^2 b_{134} (b_{72}/2+b_{183} q_2^{in}
- B_{12}^2 b_{134} b_{183} q_1^{in}/3) \right]; \nonumber \\
q_2^{fin} & = & q_2^{in}-B_{12}^2 b_{134} q_1^{in} ; \nonumber \\
p_2^{fin}  &=&  p_2^{in}+ B_{12}^2 \left[ b_{72}+ b_{128} q_1^{in}
+ 2 b_{183} q_2^{in}- B_{12}^2 b_{134} b_{183} q_1^{in} \right]; \\
q_3^{fin} & = & q_3^{in}-2 A_{12} B_{12}; \ \ \
p_3^{fin}  =  p_3^{in}+ 2 A_{12} B_{12}, \nonumber
\eeq
where
\beq
A_{12} & = & \left( q_1^{in}\,b_{47} + q_2^{in}\,b_{72} + (q_1^{in})^2\,b_{103}
+ q_1^{in}\,q_2^{in}\,b_{128} +
     p_2^{in}\,q_1^{in}\,b_{134} + \right. \nonumber \\
&& \left. (q_2^{in})^2\,b_{183} \right) \\
B_{12} & = & (q_3^{in}+p_3^{in}). \nonumber
\eeq

The action of $\exp(:h_{13}:)$ is given as follows:
\beq
q_1^{fin} & = & q_1^{in}+B_{13}^2 \left[ -b_{62}-2 b_{153} p_1^{in} -
b_{163} q_2^{in} - b_{169} p_2^{in} + \right. \nonumber \\
&& \left. B_{13}^2 b_{163} (b_{78}/2+b_{199} p_2^{in}
+ B_{13}^2 b_{163} b_{199} p_1^{in}/3) \right]; \ \ \
p_1^{fin}  =  p_1^{in}; \nonumber \\
q_2^{fin} & = & q_2^{in}-B_{13}^2 \left[ b_{78}+ b_{169} p_1^{in}
+ 2 b_{199} p_2^{in} + B_{13}^2 b_{163} b_{199} p_1^{in} \right]; \nonumber \\
p_2^{fin}  &=&  p_2^{in}+B_{13}^2 b_{163} p_1^{in} \\
q_3^{fin} & = & q_3^{in}-2 A_{13} B_{13}; \ \ \
p_3^{fin}  =  p_3^{in}+2 A_{13} B_{13}, \nonumber
\eeq
where
\beq
A_{13} & = & \left( p_1^{in}\,b_{62} + p_2^{in}\,b_{78} + (p_1^{in})^2\,b_{153}
+ p_1^{in}\,q_2^{in}\,b_{163} +
     p_1^{in}\,p_2^{in}\,b_{169} + \right. \nonumber \\
&& \left. (p_2^{in})^2\,b_{199} \right), \\
B_{13} & = & (q_3^{in}+p_3^{in}). \nonumber
\eeq

The action of $\exp(:h_{15}:)$ is given as follows:
\beq
q_1^{fin} & = & q_1^{in}-
\frac{\partial h_{15}}{\partial p_1}(p_1^{in},q_2^{in},q_3^{in});
\ \ \ p_1^{fin} = p_1^{in}, \nonumber \\
q_2^{fin} & = & q_2^{in}; \ \ \
p_2^{fin} = p_2^{in}+
\frac{\partial h_{15}}{\partial q_2}(p_1^{in},q_2^{in},q_3^{in}), \\
q_3^{fin} & = & q_3^{in}; \ \ \
p_3^{fin} = p_3^{in}+
\frac{\partial h_{15}}{\partial q_3}(p_1^{in},q_2^{in},q_3^{in}). \nonumber
\eeq

The action of $\exp(:h_{16}:)$ is given as follows:
\beq
q_1^{fin} & = & q_1^{in}-
\frac{\partial h_{16}}{\partial p_1}(p_1^{in},q_2^{in},p_3^{in});
\ \ \ p_1^{fin} = p_1^{in}, \nonumber \\
q_2^{fin} & = & q_2^{in}; \ \ \
p_2^{fin} = p_2^{in}+
\frac{\partial h_{16}}{\partial q_2}(p_1^{in},q_2^{in},p_3^{in}), \\
q_3^{fin} & = & q_3^{in}-
\frac{\partial h_{16}}{\partial p_3}(p_1^{in},q_2^{in},p_3^{in}); \ \ \
p_3^{fin} = p_3^{in}. \nonumber
\eeq

The action of $\exp(:h_{17}:)$ is given as follows:
\beq
q_1^{fin} & = & q_1^{in}; \ \ \ p_1^{fin} = p_1^{in}+
\frac{\partial h_{17}}{\partial q_1}(q_1^{in},p_2^{in},q_3^{in}), \nonumber \\
q_2^{fin} & = & q_2^{in}-
\frac{\partial h_{17}}{\partial p_2}(q_1^{in},p_2^{in},q_3^{in}), \ \ \
p_2^{fin} = p_2^{in}, \\
q_3^{fin} & = & q_3^{in}, \ \ \
p_3^{fin} = p_3^{in}+
\frac{\partial h_{17}}{\partial q_3}(q_1^{in},p_2^{in},q_3^{in}),. \nonumber
\eeq

The action of $\exp(:h_{18}:)$ is given as follows:
\beq
q_1^{fin} & = & q_1^{in}-
\frac{\partial h_{18}}{\partial p_1}(p_1^{in},p_2^{in},q_3^{in});
\ \ \ p_1^{fin} = p_1^{in}, \nonumber \\
q_2^{fin} & = & q_2^{in}-
\frac{\partial h_{18}}{\partial p_2}(p_1^{in},p_2^{in},q_3^{in}), \ \ \
p_2^{fin} = p_2^{in}, \\
q_3^{fin} & = & q_3^{in}, \ \ \
p_3^{fin} = p_3^{in}+
\frac{\partial h_{18}}{\partial q_3}(p_1^{in},p_2^{in},q_3^{in}). \nonumber
\eeq

The action of $\exp(:h_{19}:)$ is given as follows:
\beq
q_1^{fin} & = & q_1^{in};
\ \ \ p_1^{fin} = p_1^{in}+
\frac{\partial h_{19}}{\partial q_1}(q_1^{in},q_2^{in},p_3^{in}), \nonumber \\
q_2^{fin} & = & q_2^{in}; \ \ \
p_2^{fin} = p_2^{in}+
\frac{\partial h_{19}}{\partial q_2}(q_1^{in},q_2^{in},p_3^{in}), \\
q_3^{fin} & = & q_3^{in}-
\frac{\partial h_{19}}{\partial p_3}(q_1^{in},q_2^{in},p_3^{in}); \ \ \
p_3^{fin} = p_3^{in}. \nonumber
\eeq

The action of $\exp(:h_{20}:)$ is given as follows:
\beq
q_1^{fin} & = & q_1^{in}-
\frac{\partial h_{20}}{\partial p_1}(p_1^{in},p_2^{in},p_3^{in});
\ \ \ p_1^{fin} = p_1^{in}, \nonumber \\
q_2^{fin} & = & q_2^{in}-
\frac{\partial h_{20}}{\partial p_2}(p_1^{in},p_2^{in},p_3^{in}), \ \ \
p_2^{fin} = p_2^{in}, \\
q_3^{fin} & = & q_3^{in}-
\frac{\partial h_{20}}{\partial p_3}(p_1^{in},p_2^{in},p_3^{in}), \ \ \
p_3^{fin} = p_3^{in}. \nonumber
\eeq

The action of $\exp(:h_{22}:)$ is given as follows:
\beq
q_1^{fin} & = & q_1^{in}+ 3 A_{22}^2; \ \ \
p_1^{fin}  =  p_1^{in}-3 A_{22}^2 (1-2 b_{85} q_1^{in}- 3 A_{22}^2 b_{85}); \nonumber \\
q_2^{fin} & = & q_2^{in}+ 3 B_{22}^2; \ \ \
p_2^{fin}  =  p_2^{in}-3 B_{22}^2 (1-2 b_{176} q_2^{in}- 3 B_{22}^2 b_{176}); \\
q_3^{fin} & = & q_3^{in}+ 3 C_{22}^2; \ \ \
p_3^{fin}  =  p_3^{in}-3 C_{22}^2 (1-2 b_{206} q_3^{in}- 3 C_{22}^2 b_{206}), \nonumber
\eeq
where
\beq
A_{22} & = &
= {\left( -p_1^{in} - q_1^{in} + (q_1^{in})^2\,b_{85} \right) }, \nonumber \\
B_{22} & = &
{\left( -p_2^{in} - q_2^{in} + (q_2^{in})^2\,b_{176} \right) } \\
C_{22} & = &
{\left( -p_3^{in} - q_3^{in} + (q_3^{in})^2\,b_{206} \right) }. \nonumber
\eeq

\nonumsection{References}


\noindent

\begin{thebibliography}{99}
\bibitem{sanzserna} J. M. Sanz-Serna and M. P. Calvo, {\em Numerical
Hamiltonian Problems}, (Chapman \& Hall, London, 1994).
\bibitem{ruth} R. D. Ruth, A canonical integration technique,
{\em IEEE Trans. Nucl. Sci.} {\bf 30}, 2669 (1983).
\bibitem{kang} F. Kang, On difference schemes and symplectic geometry,
in: Feng Kang ed., {\em Proc. 1984 Beijing Symposium on Differential
Geometry and Differential Equations} (Science Press,
Beijing, 1985), p. 42--58.
\bibitem{lasagni} F. Lasagni, Canonical R-K methods, {\em ZAMP} {\bf 39}, 952 (1988).
\bibitem{neri} F. Neri, Department of Physics Technical Report,
University of Maryland, 1988.
\bibitem{irwin} J. Irwin, A multi-kick factorization algorithm for nonlinear
symplectic maps, {\em SSC Report No. 228}, 1989.
\bibitem{suris} Y. B. Suris, The canonicity of mappings generated by
R-K type methods when integrating the systems $x''=-\partial U/\partial x$,
{\em U.S.S.R. Comp. Maths. Math. Phys.} {\bf 29}, 149 (1989).
\bibitem{yoshida} H. Yoshida, Construction of higher order symplectic integrators,
{\em Phys. Lett. A} {\bf 150}, 262 (1990).
\bibitem{forest} E. Forest and R. Ruth, Fourth order symplectic integration,
{\em Physica D} {\bf 43}, 105 (1990).
\bibitem{channell} P. J. Channel and C. Scovel, Symplectic integration
of Hamiltonian systems, {\em Nonlinearity} {\bf 3}, 231 (1990).
\bibitem{gr1} G. Rangarajan, Invariants for symplectic maps and symplectic
completion of symplectic jets, Ph. D. thesis, University of Maryland, 1990.
\bibitem{gr2} G. Rangarajan, A. J. Dragt and F. Neri, Solvable map
representation of a nonlinear symplectic map, {\em Part. Accel.}
{\bf 28}, 119 (1990).
\bibitem{dragt1} A. J. Dragt, I. M. Gjaja, and G. Rangarajan,
Kick factorization of symplectic maps, {\em Proc. 1991 IEEE Part. Accel.
Conf.} 1621 (1991).
\bibitem{iserles} A. Iserles, Efficient R-K methods for Hamiltonian
systems, {\em Bull. Greek Math. Soc.} {\bf 32}, 3 (1991).
\bibitem{okubnor} D. Okubnor and R. D. Skeel, An explicit Runge-Kutta-Nystrom
method is canonical if and only if its adjoint is explicit, {\em SIAM J. Numer.
Anal.} {\bf 29}, 521 (1992).
\bibitem{sanzserna2} J. M. Sanz-Serna, Symplectic integrators for Hamiltonian
problems: An overview, {\em Acta Numerica} {\bf 1}, 243 (1992).
\bibitem{dragt2} A. J. Dragt and D. T. Abell, Jolt factorization of
symplectic maps, {\em Int. J. Mod. Phys. A
(Proc. Suppl.)} {\bf 2B}, 1019 (1993).
\bibitem{gjaja} I. Gjaja, Monomial factorisation of symplectic
maps, {\em Part. Accel.} {\bf 43}, 133 (1994).
\bibitem{gr3} G. Rangarajan, Symplectic completion of symplectic jets,
{\em J. Math. Phys.} {\bf 37}, 4514 (1996).
\bibitem{gr4} G. Rangarajan, Jolt factorization of the pendulum map,
{\em J. Phys. A: Math. and Gen.} {\bf 31} 3649 (1998).
\bibitem{gr5} G. Rangarajan and M. Sachidanand, Symplectic integration using solvable maps,
{\em J. Phys. A: Math. and Gen.} {\bf 33}, 131 (2000).
\bibitem{Dragt1} A. J. Dragt, Lectures on nonlinear orbit dynamics,
in: R. A. Carrigan, F. R. Huson, and M. Month eds., {\em Physics of High Energy Particle Accelerators},
AIP Conference Proceedings No. 87 (American Institute of Physics,
New York, 1982), pp. 147--313.
\bibitem{Dragt2} A. J. Dragt, F. Neri, G. Rangarajan,
D. R. Douglas, L. M. Healy and R. D. Ryne, Lie algebraic treatment of linear
and nonlinear beam dynamics, {\em Ann.
Rev. Nucl. Part. Sci.} {\bf 38}, 455 (1988)  and references therein.
\bibitem{yan} A. Chao, T. Sen, Y. Yan and E. Forest, {\em SSCL Report No. 459}, 1991.
\bibitem{berg} J. S. Berg, R. L. Warnock, R. D. Ruth, E. Forest,
Construction of symplectic maps for nonlinear motion of particles in
accelerators, {\em Phys. Rev. E} {\bf 49}, 722 (1994).
\bibitem{etienne} E. Forest, L. Michelotti, A. J. Dragt, J. S. Berg,
The modern approach to single particle dynamics for circular rings,
in: M. Month, A. Ruggiero and W. Weng, eds.,
{\em Stability of Particle Motion in Storage Rings}, AIP Conf. Proc. No. 292, (AIP,
New York, 1994).
\bibitem{Finn} A. J. Dragt and J. M. Finn, Lie series and invariant
functions for analytic symplectic maps, {\em J. Math. Phys.} {\bf 17}, 2215 (1976).
\bibitem{Cornwell} J. F. Cornwell, {\em Group Theory in Physics}, Vol.
2  (Academic Press, London, 1984).
\bibitem{Goldstein} H. Goldstein, {\em Classical Mechanics}, 2nd ed., (Addison-Wesley,
Reading, 1980).
\bibitem{giorgilli} A. Giorgilli, A computer program for integrals of motion,
{\em Comp. Phys. Comm.} {\bf 16}, 331 (1979).
\bibitem{grpoly} G. Rangarajan, Symplectic integration of Hamiltonian systems using
polynomial maps, {\em Physics Letters A} {\bf 286}, 141 (2001).
\end{thebibliography}
\end{document}